\documentstyle[psfig,times]{mn}
\def\H0{{\it H}$_0$}
\def\Ms{{\it M}$_\odot$}

\def\q0{{\it q}$_0$}

\def\Ms{{\it M}$_\odot$}

\def\nH{$N_{\rm H}$\thinspace}
\def\psqcm{cm$^{-2}$}
\def\ergpspsqcm{erg~cm$^{-2}$~s$^{-1}$}

\def\cps{ct\thinspace s$^{-1}$}
\def\Rin{$R_{\rm in}$}
\def\Rout{$R_{\rm out}$}

\def\rg{$r_{\rm g}$}
\def\phpspsqcm{ph\thinspace s$^{-1}$\thinspace cm$^{-2}$}

\def\approxlt{\raisebox{-0.6ex}{$\stackrel{<}{\sim}$}}

\title[Fe K line profile of MCG--6-30-15]
{Variation of the broad X-ray iron line in MCG--6-30-15 during a flare}
\author[K. Iwasawa et al]
{\parbox[]{6.5in} {K. Iwasawa$^1$, A.C. Fabian$^1$, A.J. Young$^1$, 
H. Inoue$^2$ and C. Matsumoto$^2$}\\
\\
$^1$ Institute of Astronomy, Madingley Road, Cambridge CB3 0HA\\
$^2$ Institute of Space and Aeronautical Science, Sagamihara, 
Kanagawa 229-8510, Japan
}
\date{}

\begin{document}

\maketitle

\begin{abstract}
We report results on the broad iron emission line of the Seyfert
galaxy MCG--6-30-15, obtained from the second long ASCA observation in
1997. The time-averaged profile of the broad line is very similar to
that seen with ASCA in 1994, so confirming the detailed model fit then
obtained. A bright flare is seen in the light curve, during which the
continuum was soft. At that time the emission line peaks around 5 keV
and most of its emission is shifted below 6 keV with no component
detected at 6.4 keV ($EW<60$ eV). This can be interpreted as the
result of an extraordinarily large gravitational redshift due to a
dominant flare occurring very close to the black hole at a radius of
$\approxlt 5$\rg.

\end{abstract}


\section{Introduction}

A major discovery from ASCA has been the discovery of a clear, broad, skewed
iron line in the spectrum of the Seyfert galaxy MCG--6-30-15 (Tanaka et al
1995). This emission line has a profile matching that expected from the
inner regions, from about 6 to 40 gravitational radii (i.e. 6 -- 40
$GM/c^2$), of a disk around a black hole (Fabian et al 1989). No simple
alternative model is capable of explaining this profile (Fabian et al 1995).
Similar skewed lines have since been found in the spectra of many other
Seyfert galaxies (Nandra et al 1997; Reynolds 1997). The broad line in
MCG--6-30-15 has also been clearly detected with BeppoSAX (Guainazzi et al
1999).

The 1994 ASCA observation of MCG--6-30-15 reported by Tanaka et al (1995)
remains however the best example of a broad line due to the good spectral
resolution of the detectors used and the long integration time of 4.5 days.
Here we report on a similar long ASCA observation of the object made in
1997. We confirm in detail the time-averaged line shape, which only shows a
small change in the `blue' horn.

During the 1994 ASCA observation the light curve of the source showed
both a flare and a deep minimum (Iwasawa et al 1996). The line profile was
seen to alter, being mostly a blue horn during the flare and then
showing only an extreme red horn during the minimum. These changes
were assumed to be due to changes in the location in the most active
regions irradiating the disk (and so producing the iron line), the
flare being on the approaching side of the disk and the minimum
emission from within the innermost stable orbit of a non-spinning
Schwarzschild black hole (Iwasawa et al 1996). This last possibility
has been explored further by Dabrowski et al (1997), Reynolds \&
Begelman (1998); Weaver \& Yaqoob (1998) and Young, Ross \& Fabian
(1998).

The light curve of the source during 1997 also shows flares and dips. We
examine in detail here the major flare seen during which both the continuum
and line show large changes. 

\section{Observations and data reduction}

MCG--6-30-15 was observed with ASCA from 1997 August 3 to 1997 August 10
with a half-day gap in the middle. It was also observed simultaneously with
Rossi X-ray Timing Explorer (RXTE) (Lee et al 1999).

The Solid state Imaging Spectrometer (SIS; S0 and S1)
was operated in Faint mode throughout the observation, using the
standard CCD chips (S0C1 and S1C3). The Gas Imaging Spectrometer (GIS; G2
and G3) was operated in PH mode. We present results mainly from the SIS data
in this Letter.
The ASCA S0 light curve in the 
0.6--10 keV band is shown in Fig. 1. 


\begin{figure}
\centerline{\psfig{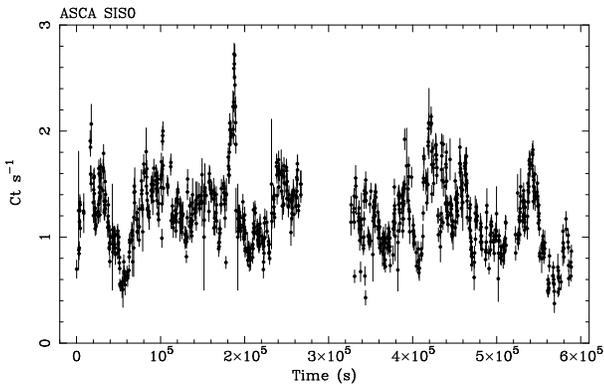}}
\caption{The ASCA SIS0 0.6--10 keV light curve of MCG--6-30-15 in August 1997.
The epoch of the light curve is 1997 August 3, 22:06:11 (UT). 
Each data point has a 128-s (or less) exposure time.}
\end{figure}

Data reduction was carried out using FTOOLS version 4.0 and 4.1 with
standard calibration provided by the ASCA Guest Observer Facility (GOF). 
The good exposure time is approximately 231 ks from each SIS
detector. The source counts are
collected from a region centred at the X-ray peak within 4 arcmin in radius
for the SIS and 5 arcmin for the GIS. The background data are taken from a
(nearly) source-free region in the same detector with the same observing
time.
The efficiency of the S1 detector below 1 keV 
appears to be severely reduced due to
the Residual Darkframe Distribution (RDD, Dotani 1998), which the 
current response matrix (generated from calibratione 
files in the FTOOLS version 4.1 release) does not take into account for.
The RDD effect on the S0 data from 1CCD observations 
has been found to be very little 
(Dotani 1998). Therefore, the S1 data below 1 keV were discarded 
for the spectral analysis presented here.
The energy resolution of the SIS at 6.4 keV when
the observation was  carried out had degraded to 
$\sim 250$ eV (FWHM), about twice that
attained immediately after launch of the satellite.

\section{Comparisons with the 1994 long observation}

The average count rates in the 0.6--10 keV band from the S0/S1 detectors are
1.16/0.93 \cps (cf. 1.53/1.25 \cps\ during the previous long observation
in 1994).
The average observed fluxes are $1.47\times 10^{-11}$\ergpspsqcm\ in the 
0.5--2 keV band and $3.41\times 10^{-11}$\ergpspsqcm\ in the
2--10 keV band.

\subsection{Total energy spectrum and warm absorber}

The observed 0.6--10 keV X-ray flux during the present observation is
lower by 26 per cent than that during the 94 long observation. The
3--10 keV (the iron K band, 4--7.5 keV, excluded) power-law slope is
$\Gamma = 1.94^{+0.06}_{-0.07}$, which is similar to the 94 data.
Features of the warm absorber detected are two edges due to OVII at
0.72 keV and OVIII at 0.85 keV (e.g., Otani et al 1996) and one at 1.1
keV, probably due to NeIX and/or Fe L. The 97 spectrum is harder in the
low energy band than the 94 spectrum, which may be explained by an
increase in absorption. Details will be reported by Matsumoto et al
(in prep).

\subsection{The iron K line}

The 3--10 keV data were investigated for iron K line emission. The
continuum spectrum was modelled with a power-law reflection model ({\tt
pexrav}, Magdziarz \& Zdziarski 1995) modified by cold absorption of
\nH $= 7\times 10^{20}$\psqcm (which has virtually no effect on the
3--10 keV continuum). The parameters of {\tt pexrav}, apart from
photon index and normalization, were matched to the previous
measurements of MCG--6-30-15: the cut-off energy, 130 keV (from the
BeppoSAX observation by Guainazzi et al 1999); reflection intensity,
corresponding to $\Omega/2\pi =1$ (Guainazzi et al 1999; Lee et al
1999); iron abundance of unity (Lee et al 1999); and inclination
of the reflecting slab, 30$^{\circ}$ (Tanaka et al 1995).
The line feature is fitted by the
diskline model for a Schwarzschild black hole (Fabian et al 1989). The rest
energy of the line emission is assumed to be 6.4 keV, appropriate for cold
iron. A power-law ($\Gamma = 1.96^{+0.04}_{-0.03}$) 
modified by reflection plus a diskline model provide
a good fit ($\chi^2 = 724.7$ for 729 degrees of freedom).
The best-fit parameters of the diskline model are shown in Table 1.


\begin{figure}
\centerline{\psfig{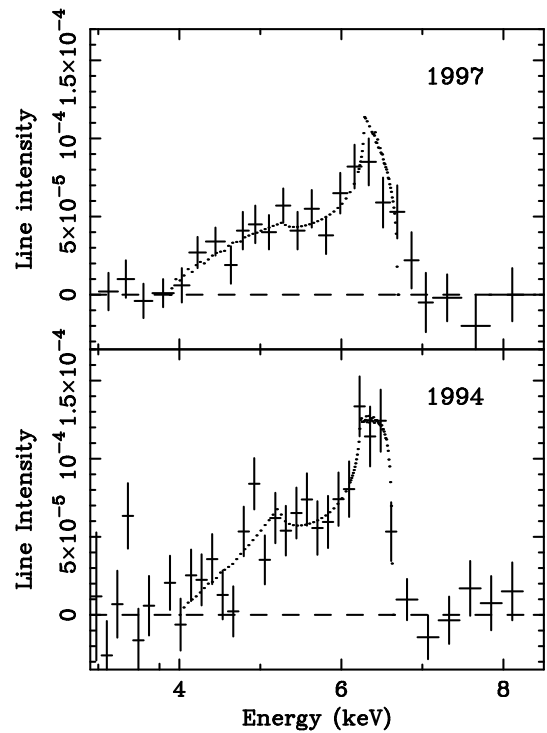}}
\caption{The iron K emission-line profiles of MCG--6-30-15 obtained
from the long observations in 1997 (upper panel) and 1994 (lower
panel from Tanaka et al 1995). Dotted line shows best-fit diskline model
for a Schwarzschild black hole. See Table 1 for the diskline parameters
for the 97 profile.}
\end{figure}


\begin{table*}
\begin{center}
\caption{The best-fit parameters of the diskline fit to the
averaged Fe K line profile. The continuum is modelled by a power-law modified
by reflection ({\tt pexrav}, see text).
The diskline model by Fabian et al (1989) is
used. (1) The rest energy of the line emission ;
(2) power-law index of the radial emissivity law ($\propto r^{-\alpha}$);
(3),(4)Inner and outer radii of the line emitting disk in a unit of 
gravitational radius (\rg);
(5) inclination angle of the accretion disk; (6) efficiency-corrected line 
intensity; (7) equivalent width of the line emission calculated assuming 
all the line emission is concentrated in a narrow energy range around 6.4 keV.}
\begin{tabular}{ccccccc}
(1) & (2) & (3) & (4) & (5) & (6) & (7) \\
$E$ & $\alpha $ & \Rin & \Rout & $i$ & $I$ & $EW$ \\
keV & & \rg & \rg & deg & \phpspsqcm & eV \\[5pt]
6.4 & $4.1^{+2.0}_{-2.0}$ & $6.7^{+0.9}_{-0.7}$ & $24_{-10}^{+20}$ & 
$32^{+2}_{-2}$ & $1.36^{+0.26}_{-0.24}\times 10^{-4}$ & $388^{+74}_{-68}$ \\
\end{tabular}
\end{center}
\end{table*}

The efficiency-corrected line profile\footnote{This is not an `unfolded'
spectrum but obtained from the ratio, the data divided by the power-law model
(folded through the detector response) best-fitting the neighbouring 
continuum, multiplied by the power-law (in original form). The plot is 
therefore independent from the model used for fitting the line.} 
for the present data set is 
shown in Fig. 2, along with the one from the previous long observation
in 1994 (Tanaka et al 1995). 
The profile from the present observation
appears to be less bright in the blue peak while it shows 
a slightly more extended red wing. During the previous observation, the
bright flare ({\it i-3}) data showed a narrow-core-dominated line profile
(Iwasawa et al 1996).
Such a line shape is not found during the present observation.
However, this is not sufficient to explain the difference in the blue horn 
of the time-averaged line profiles between the two observations.
Although the overall line shape is similar between 94 and 97,
the steeper radial emissivity index suggests that the mean weight of the
line emissivity may be slightly shifted towards the inner part of 
the accretion disk in 97 as compared to 94.

\section{The major flare}

Changes in the iron line profile were investigated in time sequence,
details of which will be reported elsewhere.
Here we show the peculiar behaviour of the energy spectrum and iron line
during the 
major flare, which occured around $1.9\times 10^5$ s in the light 
curve (see Fig. 1 and Fig. 3 for a detailed version of the light curve
around the flare).
The continuum is steeper than usual ($\Delta\Gamma \sim 0.17$), 
particularly at low energies (1--3 keV band), as shown in Fig. 4.


\begin{figure}
\centerline{\psfig{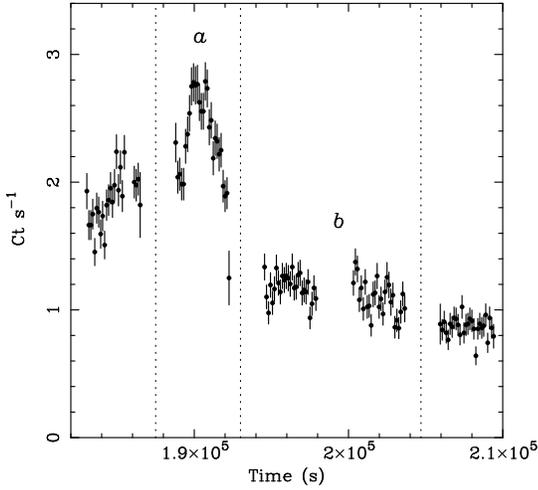}}
\caption{The ASCA SIS0 light curve of MCG--6-30-15 around the brightest
flare. The epoch of the light curve is 1997 August 3, 22:06:11.
The iron K line profiles during the time intereval {\it a} and {\it b}
are shown in Fig. 5.}
\end{figure}


\begin{figure}
\centerline{\psfig{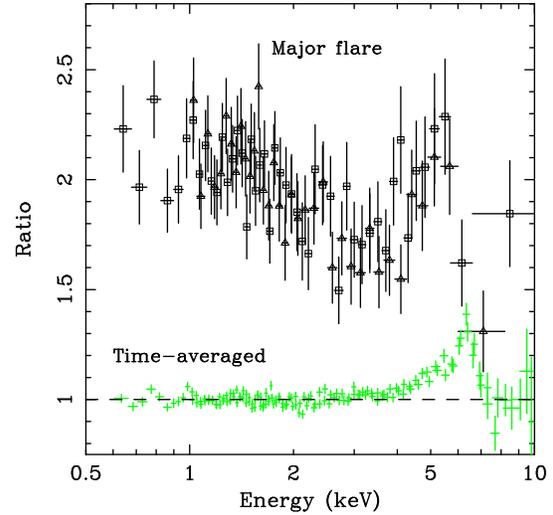}}
\caption{The major flare spectrum (taken from the time-intereval {\it a}
in Fig. 3, squares: S0; triangles: S1) divided by
the best-fit model for the time-averaged spectrum. The same plot for
the time-averaged spectrum is also shown. The model is 
a power-law ($\Gamma = 1.94$) modified by three absorption edges at
0.71, 0.85, and 1.1 keV, and cold absorption (\nH $=9\times 10^{20}$\psqcm).
The Fe K band was excluded from the fit so that the Fe K line features 
remain in the plot.}
\end{figure}


\begin{figure}
\centerline{\psfig{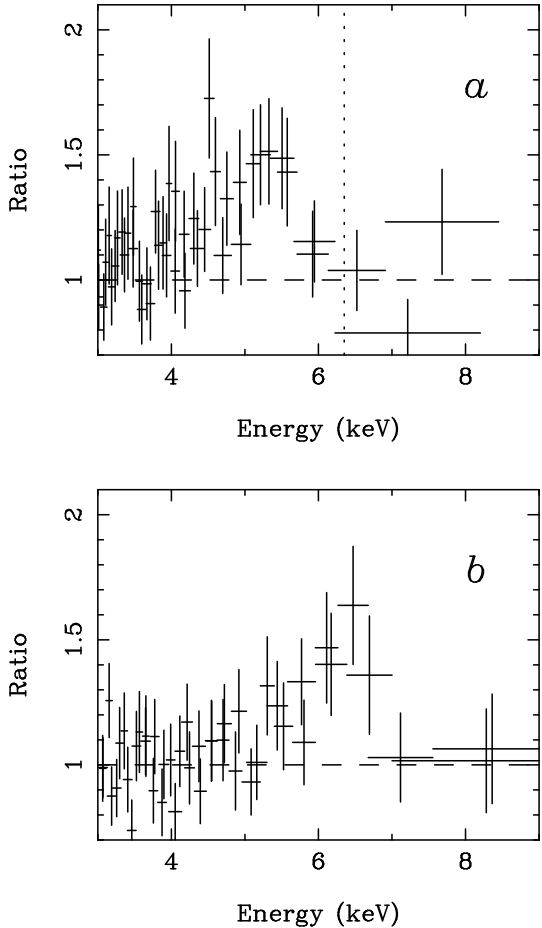}}
\caption{The ratio plot of the SIS data to the baseline power-law models
for the time intervals {\it a} and {\it b} in Fig. 3. 
The vertical dotted line in the upper panel
indicates the rest energy of the cold iron K$\alpha$ emission line,
6.4 keV. Almost all of the line emission is redshifted well below the rest
line energy during the flare ({\it a}). When the flare ceased ({\it b}), 
the line shape recovered to the ordinary one, as shown in the lower panel.}
\end{figure}

\subsection{Iron K line emission with large redshift?}

Excess emission above a power-law continuum, which is 
presumably due to broad iron K line emission, is found
in the 4--7 keV range. 
The ratio plots of the data and the baseline power-law
model for the intervals of the flare peak ({\it a}) and of
the subsequent dip ({\it b}) are shown in Fig. 5.
The baseline model is obtained by fitting a power-law
to the data adjacent to the iron line band.

The line profile of the flare peak ({\it a}) shows a sharp decline at
$\sim $5.6 keV, which is far below the rest energy of the line
emission of Fe~K$\alpha$, 6.4 keV, and a red-wing extending down to
$\sim $3.5 keV. We have checked the GIS data which confirm the SIS
result. No line emission is detected at 6.4 keV ($<3\times
10^{-5}$\phpspsqcm, $EW<60$ eV, 90 per cent upper limits obtained from
the joint fit to the SIS and GIS data). This extremely redshifted line
profile cannot be explained by the diskline for a Schwarzschild black
hole because of insufficient gravitational redshift (it may arise from
infalling gas, as proposed by Reynolds \& Begelman 1997, but then
there would also be a large absorption edge; Young et al 1998). On the
other hand, the diskline model for a Kerr black hole by Laor (1991;
which is for a maximally-rotating black hole) gives a good fit. The
result of this diskline fit is shown in Table 5, where the rest line
energy, the inner radius and inclination of the disk are assumed to be
6.4 keV, 1.235\rg, and 30$^{\circ}$. The outer radius is constrained
well at $(5\pm 1)$\rg\ due to the well-defined decline of the
redshifted, blue peak. The inferred negative emissivity index (it is
poorly constrained) and the well-constrained outer radius perhaps
suggest the line emission is concentrated in annuli around $\sim$
5\rg. A fit with a double-gaussian to the line profile is also given
in Table 5. The line intensity of the line is about 3 times larger
than that of the time-averaged one. The EW is $\sim 700$ eV when
computed with respect to the continuum at 6.4 keV but $\sim 400$ eV to
the continuum in the energy range of the observed line.

The line profile models available in XSPEC, the fits of which have
just been reported above, are for complete disk annuli. It is possible
however that just part of an annulus of the disk is irradiated during
a flare, i.e. that part immediately below the flare itself. This
offers further possible locations for the flare such as on the
receding side of the disk where the peak at 5 keV is mostly due to the
doppler effect, or on the approaching side much closer in where it is
due to gravitational redshift. We show in Fig. 6 the locus of points
which cause a 6.4 keV line in the disk frame to appear at 5 keV for an
observer seeing the disk around a maximally spinning black hole at an
inclination of $30^{\circ}$. We have also fit the flare spectrum with
model line profiles created from a disk divided into 36 azimuthal
sectors and 24 radial bins between 1 and 25
\rg. Although acceptable fits at $\sim 3\sigma$ level
are found for most of the points around this locus, the region at
small radii ($\sim 2.5$\rg) on the approaching side of the disk is
favoured most (see the confidence contours in Fig. 6). This is mainly
due to the excess flux around 4~keV shown in Fig~5a and not the choice
of grid size.

In the subsequent time-interval {\it b}, the averaged 0.6--10 keV
count rate dropped by a factor of 2.2, compared with the flare interval
(see Fig. 3). The line shape has then recovered to the ordinary one, as
seen in the time-averaged spectrum (Fig. 5b). The line intensity,
$(2.2\pm 1.3)\times 10^{-4}$\phpspsqcm,
also dropped by a factor of 2.

\begin{figure}
\vspace{-5mm}
\centerline{\psfig{figure=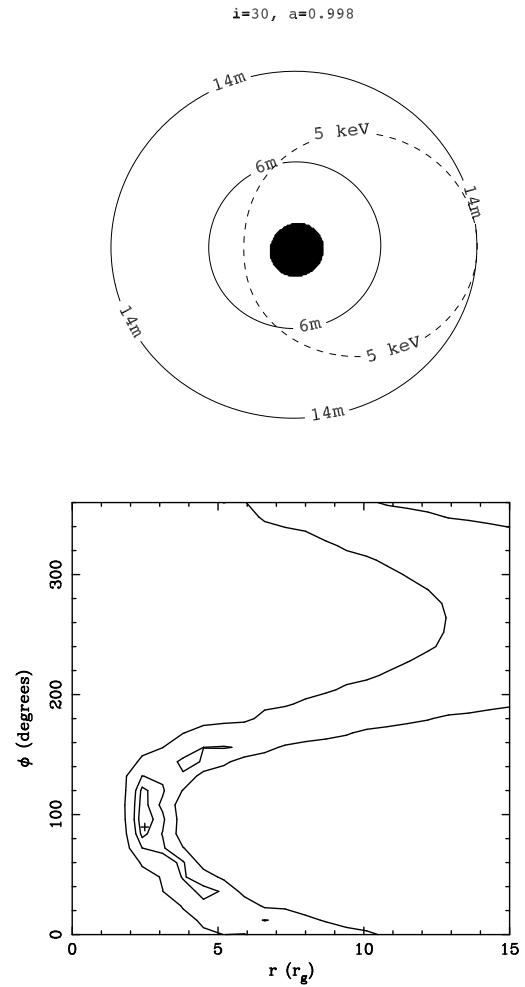,width=0.4\textwidth}}
\vspace{-5mm}
\centerline{\psfig{figure=fig6b.ps,width=0.38\textwidth,angle=270}}
\caption{Upper panel: Locus of points which shift a 6.4keV line to 5~keV.
The accretion disk around a maximally spinning (angular momentum,
a = 0.998) black hole viewed at an inclination angle of $30^{\circ}$ is 
assumed. The approaching side of the disk is on the left (the receding side 
on the right).
Lower panel: Confidence contours in the radius and azimuthal-angle plane
obtained on fitting model lines from
separate parts of a disk to the line profile data shown in Fig. 5a. Azimuthal
angle $\phi$ is measured clock wise from the near side of the disk.
Contours are drawn at 68, 90 and 99 per cent confidence levels for two 
parameters of interest. The possibility that a whole ring at $\sim 5$\rg\ 
yields an acceptable fit (see text) is not evident from this plot, since
if it involves only single small azimuthal sector at each radius.}
\end{figure}


\begin{table*}
\begin{center}
\caption{Diskline and double-gaussian fits to the line profile from the
flare interval ({\it a} in Fig. 3). The diskline model is for 
a maximally rotating Kerr hole
by Laor (1991). The emissivity index is defined by a power-law function
of radius ($\propto r^{-\alpha}$).}
\begin{tabular}{cccccccc}
\multicolumn{7}{c}{Diskline (Laor 1991)} \\[5pt]
$E$ & $\alpha $ & \Rin & \Rout & $i$ & $I$ & $\chi^2$/dof \\
keV & & \rg & \rg & deg & $10^{-4}$\phpspsqcm & \\[5pt]
6.4 & $-4.5(<2.7)$ & 1.235 & $4.6^{+0.6}_{-0.4}$ & 30 & $4.9^{+3.0}_{-2.2}$ &
102.0/143 \\[8pt]
\multicolumn{7}{c}{Double-gaussian} \\[5pt]
$E_1$ & $\sigma_1$ & $I_1$ & $E_2$ & $\sigma_2$ & $I_2$ & $\chi^2$/dof \\
keV & keV & $10^{-4}$\phpspsqcm & keV & keV & $10^{-4}$\phpspsqcm & \\[5pt]
$5.39^{+0.16}_{-0.23}$ & $0.19(<0.44)$ & $2.0^{+2.6}_{-1.6}$ & 
$4.49^{+0.57}_{-1.30}$ & $0.51(<1.93)$ & $2.4^{+4.6}_{-2.1}$ & 101.6/140 \\
\end{tabular}
\end{center}
\end{table*}

\section{Discussion}

The long ASCA observation of 1997 has confirmed in detail the broad iron
line in the Seyfert galaxy MCG--6-30-15. The time-averaged emission appears
to originate from a disk extending between about 6 and 40 \rg\
of a massive black hole. If the rapid variability of this source is
due to flares above the accretion disk, then the long term constancy of the
line profile indicates that there are usually several flares at once on the
disk and that the distribution of flares is almost
constant in a time-averaged sense.

The spectrum of the source, and in particular the line profile, changed
dramatically during a bright flare. Unlike the bright blue horn apparent
during the 1994 flare, we now see essentially a bright red horn. 
If this is interpreted in the context of a relativistic diskline,
the dominant flare must occur at smaller radii than usual.
There are two possible locations of the flare, depending on the mass of 
the black hole in MCG--6-30-15.

One interesting possibility is a flare localized on the approaching
side of the disk at $\sim 2.5$\rg\ (see Fig. 6). The duration of the
flare is about 1 hour while the Keplerian orbital time around a
$10^7M_7$\Ms\ black hole is $10^4M_7r_1^{3/2}$s at $10r_1$\rg. If the
flare is confined within, say, 1/6 orbit at 2.5\rg\ (see Fig. 6), the
duration of the flare (and peculiar line shape) requires the black
hole mass to be larger than $20 M_7$\Ms. Therefore this solution is
valid only if the black hole in MCG--6-30-15 is more massive than
$10^8$\Ms.

The spectral fit with the model for azimuthally-averaged line emission
(Laor 1991) suggested that the line emission may be produced in a
narrow range of radii around 5\rg\ during the major flare (see Table
2). The duration of the flare corresponds to $1M_7^{-1}$ orbital time
at 5\rg. This is the preferred solution if the black hole mass is
significantly smaller than $10^8$\Ms.

There are some difficulties for the first interpretation. In order to
restrict the line production to part of the disk, the flare must be
placed very close to the disk surface. A flare on the approaching side
of the disk is generally expected to be amplified due to relativistic
beaming (e.g., Karas et al 1992), which appears to be consistent with
the observed flux variation. However, at a small radii such as 2.5\rg\
on a disk inclined at $30^{\circ}$, gravitational redshift and frame
dragging overwhelm doppler boosting so that the emission reaching a
distant observer is suppressed by more than an order of magnitude when
the X-ray source is placed at 1\rg\ above the disk (M. Ruszkowski,
priv. comm.). Therefore the flare would have to be intrinsically much
more intense than observed.  This may be possible if the emitted power
increases rapidly towards inner radii around a spinning black hole.
The strong light deflection implies that the reflection from the disk
should also be enhanced by a factor of $\sim 2$ at the same time (see
also Martocchia \& Matt 1996). The closeness of the continuum source
may also cause the disk surface to be highly ionized. Although the
high energy end of the ASCA data is rather noisy, the 6--10 keV
spectrum during the flare ($\Gamma = 1.5\pm 0.4$) suggests a possible
spectral flattening which could be due to strong reflection.

The interpretations discussed above are, of course, not unique, but
both require that the accretion disk extends close to the central
black hole and that it spins rapidly, as suggested by Iwasawa et al
(1996).

Although the profile of the broad line in MCG--6-30-15 appears to be
fairly constant in a time-averaged sense, it does undergo dramatic
changes every few days. Such changes offer interesting possibilities
with which to probe different parts of the disk and to map the
innermost regions about the black hole.

The lack of a narrow 6.4 keV line during the flare (the 90 per cent
upper limit of intensity is only 40 per cent of the blue peak
intensity of the time-averaged line) confirms the suggestion made by
the previous ASCA observation (Iwasawa et al 1996) that there is
little line emission from far out in the disk or torus.
A narrow 6.4 keV line might be delayed by an hour or so, if it
is produced around $100M_7$\rg. Any line emission from farther
out should be more constant because the variability of the line
is smeared out. Evidence for such line emission appears to be weak.

\section*{Acknowledgements}

We thank all the members of the ASCA team. ACF and KI thank Royal
Society and PPARC, respectively, for support. Chris Reynolds is
thanked for his useful comments.

\end{document}